\documentclass[preprint,showpacs,preprintnumbers,amsmath,amssymb]{revtex4-1}
\input{epsf}

\usepackage{graphicx}% Include figure files
\usepackage{dcolumn}% Align table columns on decimal point

\begin{document}

%\preprint{APS/123-QED}

\title{An iterative approach for amplitude amplification with nonorthogonal measurements}
\author{H. T. Ng${}^1$  and Franco Nori${}^{1,2}$}
\affiliation{${}^{1}$Advanced Science Institute,
RIKEN, Wako-shi, Saitama 351-0198, Japan}
\affiliation{${}^{2}$Physics Department, The University of Michigan, Ann Arbor, Michigan 48109-1040, USA
}
\date{\today}

\begin{abstract}
Using three coupled harmonic oscillators, 
we present an amplitude-amplification method
for factorization of an integer.  
We generalize the method in [arXiv:1007.4338]
by employing non-orthogonal measurements on
the harmonic oscillator.
This method can increase the probability of obtaining the factors
by repeatedly using the nonlinear
interactions between the oscillators and non-orthogonal measurements.
However, this approach requires an exponential
amount of resources for implementation. 
Thus, this method
cannot provide a speed-up over classical
algorithms unless its limitations are resolved.
\end{abstract}


\maketitle

\section{Introduction}
Quantum computing based on qubits has attracted considerable attention 
(see, e.g., \cite{Feynman,Deutsch1,Ekert,Nielsen,Kaye,Buluta1,Buluta2}).  
There are several candidates to realize quantum computers, such as using 
nuclear spins in molecules, photons, trapped ions, superconducting circuit and quantum dots (see, e.g., \cite{Buluta2,You}).  
However, it is still a great challenge to build a large-scale quantum computer.  

Quantum computers can significantly outperform classical computers 
in doing some specific tasks \cite{Ekert,Nielsen,Kaye,Childs}.  
For example, two important quantum algorithms are 
Shor's \cite{Shor} and Grover's \cite{Grover}.  
Shor's algorithm \cite{Shor} can factorize a large integer in polynomial time, offering
an exponential speed-up over classical computation.  
Grover's algorithm \cite{Grover} gives a quadratic speed-up in searching database.  
This search algorithm has been found to be very useful in other related problems \cite{Nielsen,Kaye,Childs}.  
To date, the study of quantum algorithms is a very active area of research (see, e.g., \cite{Childs}).

Using three coupled harmonic oscillators, we have recently proposed \cite{Ng} 
an alternative approach for factoring integers.  
We consider these three harmonic oscillators to be coupled together via nonlinear interactions \cite{Ng}.  
To factorize an integer $N$,  
this approach involves only three steps: initialization, time evolution, and 
conditional measurement.  In this approach, the states of the first two 
harmonic oscillators are prepared in a number-state basis, 
while the state of the third oscillator is prepared in a coherent state.  
The states of the first two harmonic oscillators encode the trial factors of the number $N$.  
The nonlinear interactions between the oscillators produce coherent states that simultaneously rotate
in phase space with different effective frequencies, which are proportional to the product of two trial 
factors \cite{Ng}.  In this way, {\it all} possible products of any two trial factors can be {\it simultaneously} computed,
and then they are ``written'' to the rotation frequencies of the coherent states in {\it a single step}.  
The resulting state of the first two oscillators is the factors' state \cite{Ng}
by performing a conditional measurement of a coherent state rotating with an effective frequency 
which is proportional to $N$.  However, the probability of obtaining this coherent state becomes low 
when $N$ is large.  In this paper, we can circumvent this limitation by using 
an iterative method for increasing 
the chance of finding the states of the factors.  This amplitude-amplification method 
involves a number of iterations,
where each iteration is very similar to the factoring approach we recently proposed \cite{Ng}.

Now we briefly describe this amplitude-amplification method for factorization using three coupled harmonic oscillators.  
Let us now consider the first step of our approach.
Initially,  the first two harmonic oscillators are in a number-state basis 
and the third oscillator is in a coherent state.  Let the three coupled harmonic oscillators evolve for a 
period of time.  The detection is then conditioned on a coherent state with a 
rotation frequency being proportional to $N$.  
The probability of finding this coherent state can be adjusted by choosing both an 
appropriate period of time evolution 
and magnitude of the coherent state.  
Here we find that this probability is not small.  
Indeed, the probability of finding the factors' state can be 
increased by a factor which is the reciprocal of the probability of obtaining this coherent state.  But the word "probability" in these sentences is different to the total probability of
obtaining the factors. 

The resulting states of the first two oscillators, after the first 
step, are used as new input states in the second step of our approach.  Also, the state of the third 
oscillator is now prepared as a coherent state with the same, or higher, 
magnitude.  By repeating the same procedure described in the first step, 
we can obtain the states of the factors with a much higher probability.  We then iterate these procedures $L\sim(\log_2{N})$ 
times, until the probability of finding the factors' state is close to one.  

As an example of how this method works, we show how to factorize 
the integer $1,030,189=1009{\times}1021$.  
Here the probabilities of obtaining coherent states, with rotation frequencies
proportional to $N$, are larger than 0.1 in each iteration.  
The probability of finding the factors can reach nearly one after 12 iterations. By comparing with the examples $N=101,617$ and 10,961, 
we can show that the required number of iterations for factoring logarithmically scales with $N$.

However, this approach requires an exponential amount
of resources for its implementation.  First, it 
requires an exponential amount of energy to encode a 
number $N$ onto the state of a harmonic oscillator.
The input energy scales with the number
$N$.  Therefore, the required energy becomes enormous 
if the number $N$ is large. 
Second, it is necessary to prepare an exponential size $O(N)$ 
of ensemble of oscillators for this iterative approach. 
This is because a large number of oscillators are abandoned 
after a conditional measurement. A number at least of 
order of O$(N)$ oscillators are thus required to be prepared for 
the input size $N$.  Therefore, this approach does not 
provide any speed-up for factorization compared to classical algorithms.

This paper is organized as follows:  In section~II, we introduce a system of coupled 
harmonic oscillators.  In section~III, we study the quantum dynamics of the coupled harmonic 
oscillators starting with a product state of number states and a coherent state.  
In section~IV,  we propose an amplitude-amplification method to factorize an integer using
three coupled harmonic oscillators.  We discuss the convergence and performance of this factoring algorithm.  
For example, we show how to factor the number $1,030,189$ using this approach.
In section~V, we discuss 
the problems and limitations of this approach.    
Finally, we make a summary in section~VI.

\begin{center}
 
\end{center}

\section{System}  
We consider a system of $(A+B)$ coupled harmonic oscillators.  The Hamiltonian of 
the $j$-th harmonic oscillator is 
written as
\begin{eqnarray}
H^{j}_{\rm osc}&=&\frac{P^2_j}{2m_j}+\frac{m_j\omega_j{X^2_j}}{2},
\end{eqnarray}
where $\omega_j$ is the frequency of the harmonic oscillator, $m_j$ is the mass of 
the particle, and $j=1,\ldots,A+B$.
The operators $X_j$ and $P_j$ are the position and momentum operators, which satisfy 
the commutation relation, $[X_j,P_j]=i\hbar$.
The annihilation and creation operators of the $j$-th harmonic oscillator are defined as
\begin{eqnarray}
\label{acoperator}
a_j&=&\sqrt{\frac{m_j\omega_j}{2\hbar}}\bigg(X_j+\frac{iP_j}{m_j\omega_j}\bigg),\\
a^\dag_j&=&\sqrt{\frac{m_j\omega_j}{2\hbar}}\bigg(X_j-\frac{iP_j}{m_j\omega_j}\bigg).
\end{eqnarray}
The operators $a_j$ and $a^\dag_j$ obey the commutator $[a_j,a^\dag_j]=1$.
The Hamiltonian of the three harmonic oscillators can be expressed in terms of 
the annihilation and creation operators:
\begin{equation}
 H_0=\hbar\sum^{A+B}_{j=1}{\omega_j}a^\dag_ja_j,
\end{equation}
Here we have ignored the constant term.  

We consider the harmonic oscillators coupled to each other via nonlinear interactions \cite{Ng}.
Such nonlinear interactions can be described by the Hamiltonian $H_I$ as
\begin{equation}
\label{nonlinearfunction}
 H_I={\hbar}\sum^{A+B}_{k=A+1}f_k(n_1,\ldots,n_{A})a^\dag_{k}a_{k}\\
\end{equation}
where $f_k(n_1,\ldots,n_{A})$ are linear or nonlinear-operator functions 
(excluding divisions) of the number operators 
$n_j=a^\dag_ja_j$, for $j=1,\ldots,{A}$, and $k=A+1,\ldots,{A+B}$.  

The total Hamiltonian $H=H_0+H_I$ can be written as
\begin{equation}
\label{Hamiltonian}
 H=\hbar\sum^{A+B}_{j=1}{\omega_j}a^\dag_ja_j+{\hbar}\sum^{A+B}_{k=A+1}f_k(n_1,\ldots,n_{A})a^\dag_ka_k.
\end{equation} 
The Hamiltonians $H_0$ and $H_I$ commute with 
each other, i.e, ${[H_0,H_I]}=0$.  
The total Hamiltonian $H$ in Eq.~(\ref{Hamiltonian}) is exactly solvable.
The eigenstate $|E_{m_1,\ldots,m_{A+B}}\rangle$ of the Hamiltonian $H$ is a product state of number states $|m_j\rangle_j$
of the harmonic oscillators, i.e.,
\begin{eqnarray}
|E_{m_1,\ldots,m_{A+B}}\rangle&=&\prod^{A+B}_{j=1}|m_j\rangle_j,
\end{eqnarray}
corresponding to an eigenvalue $E_{m_1,\ldots,m_{A+B}}$
\begin{equation}
E_{m_1,\ldots,m_{A+B}}=\sum^{A+B}_{j=1}\omega_jm_j+\sum^{A+B}_{k=A+1}f_k(m_1,\ldots,m_{A})m_k.
\end{equation}

\section{Quantum dynamics in phase space}
\begin{figure}[ht]
\centering
\includegraphics[height=7.0cm]{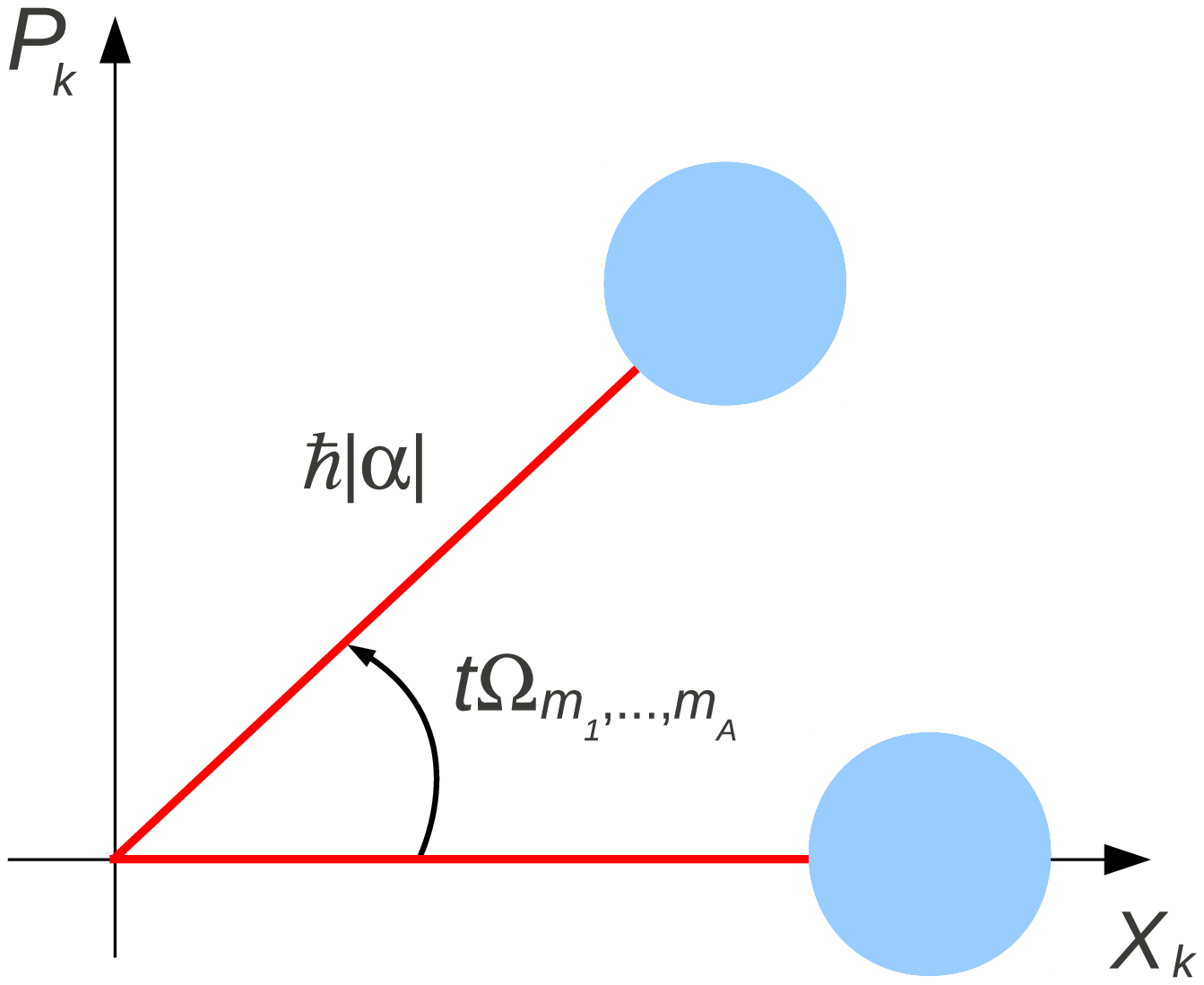}
\caption{ \label{phase_diag} (Color online) Schematic diagram of the time evolution in phase space of 
the coherent state of the harmonic oscillator $k$, corresponding to the product of 
number states $\prod_{j}|m_j\rangle_{j}$, where $j=1,\ldots,A$.   
The coherent state with complex amplitude $\alpha$ is here depicted as a light blue circle in phase space ($X_k$,$P_k$), 
where $X_k$ and $P_k$ represent the position and the momentum of the harmonic oscillator $k$.   
After a time $t$, the coherent state rotates about the origin with an angle ${t}\Omega_{m_1,\ldots,m_A}$.  
}
\end{figure}

We study the time evolution of the $k$-th harmonic oscillator in phase space starting with a state
$\prod_{j}|m_j\rangle_{j}\otimes|\alpha\rangle_{k}$, where $\prod_{j}|m_j\rangle_{j}$ is the product state of
the number states of the harmonic oscillators, and $|\alpha\rangle_{k}$ is the coherent
state of the $k$-th harmonic oscillator, for $j=1,\ldots,{A}$, and $k$ is a number from $A+1$ to $A+B$.  
By applying the time-evolution operator $U(t)=\exp(-iHt)$ [$H$ is the Hamiltonian in Eq.~(\ref{Hamiltonian})]
to the initial state $\prod_{j}|m_j\rangle_{j}|\alpha\rangle_k$,
it becomes 
\begin{eqnarray}
&&U(t)\prod_{j}|m_j\rangle_{j}|\alpha\rangle_k\nonumber\\
&=&\exp\bigg[-i\bigg(\sum_j\omega_j{m}_j\bigg)t\bigg]\prod_{j}|m_j\rangle_{j}\nonumber\\
&&\times\exp{\big\{-i[\omega_k+f_k(m_1,\ldots,m_{A})]a^\dag_ka_k{t}\big\}}|\alpha\rangle_k,\nonumber\\
\label{U_inputstate}
&=&\exp\bigg[-i\bigg(\sum_j\omega_j{m}_j\bigg)t\bigg]\prod_{j}|m_j\rangle_j|\alpha_{{m_1,\ldots,m_A}}(t)\rangle_k,
\end{eqnarray}
where $\alpha_{m_1,\ldots,m_A}(t)$  is a complex function, i.e.,
\begin{eqnarray}
%\label{alpha}
%\alpha(m_1,\ldots,m_{A},t)&=&\exp[-i\Omega(m_1,\ldots,m_{A})t]\alpha,\\
\alpha_{m_1,\ldots,m_A}(t)&=&\exp[-i\Omega_{m_1,\ldots,m_{A}}t]\alpha,
\end{eqnarray}
and $\Omega_{m_1,\ldots,m_{A}}$ is an effective rotation frequency of the coherent state in phase space
\begin{eqnarray}
\label{Omega_F_k}
\Omega_{m_1,\ldots,m_{A}}&=&\omega_k+f_k(m_1,\ldots,m_{A}).
\end{eqnarray}
In Eq.~(\ref{U_inputstate}), we have used the relation \cite{Barnett},
\begin{equation}
\exp(-i\vartheta{a^\dag_k{a}_k})|\alpha\rangle_k=|\alpha\exp(-i\vartheta)\rangle_k,
\end{equation}
where $\vartheta$ is a phase factor.

Note that the product state of the number states $\prod_{j}|m_j\rangle_{j}$ is an invariant; namely
it does not change with time.
Nonlinear interactions, described by the Hamiltonian $H_I$ in Eq.~(\ref{nonlinearfunction}), cause
the coherent state of oscillator $k$ to rotate about the origin with a 
frequency $\Omega_{m_1,\ldots,m_{A}}$ in phase space.
From  Eq.~(\ref{Omega_F_k}), the frequency $\Omega_{m_1,\ldots,m_{A}}$ depends on 
the number states $\prod_{j}|m_j\rangle_{j}$ of the $A$ oscillators.
A schematic diagram of the time evolution of the coherent state of the harmonic oscillator $k$ in phase space 
is shown in Fig.~\ref{phase_diag}.

\section{Factorization}
\begin{figure}[ht]
\centering
\includegraphics[height=4.0cm]{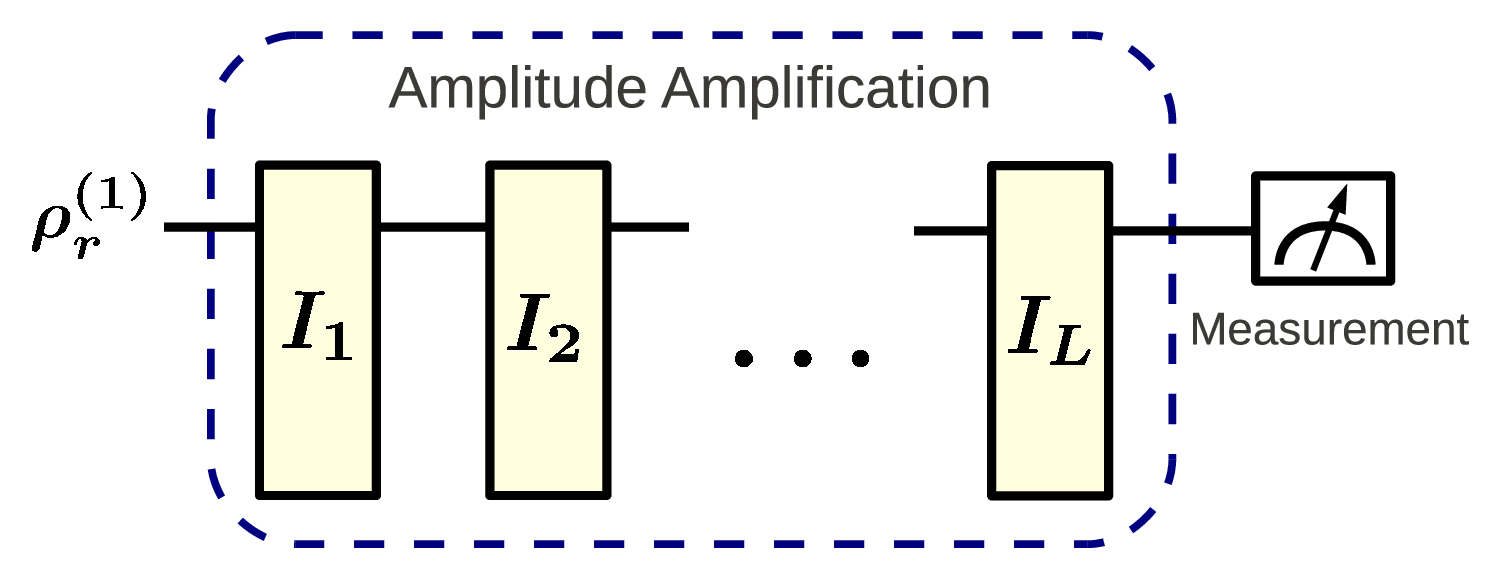}
\caption{ \label{factqcircuit} (Color online)  Factoring algorithm using three coupled harmonic oscillators.
Initially, the state $\rho^{(1)}_r$ of the harmonic oscillators 1 and 2, is prepared in a number-state basis.  
By repeatedly applying $L$ times the iterations $I_l$ (as shown in Fig.~\ref{qcir_iteration}), 
the resulting state of the harmonic oscillators 1 and 2 becomes the state of the factors of $N$.  
Finally, the factors of the number $N$ can be obtained by measuring the state of 
the oscillators 1 and 2.  
}
\end{figure}

\begin{figure}[ht]
\centering
\includegraphics[height=4.5cm]{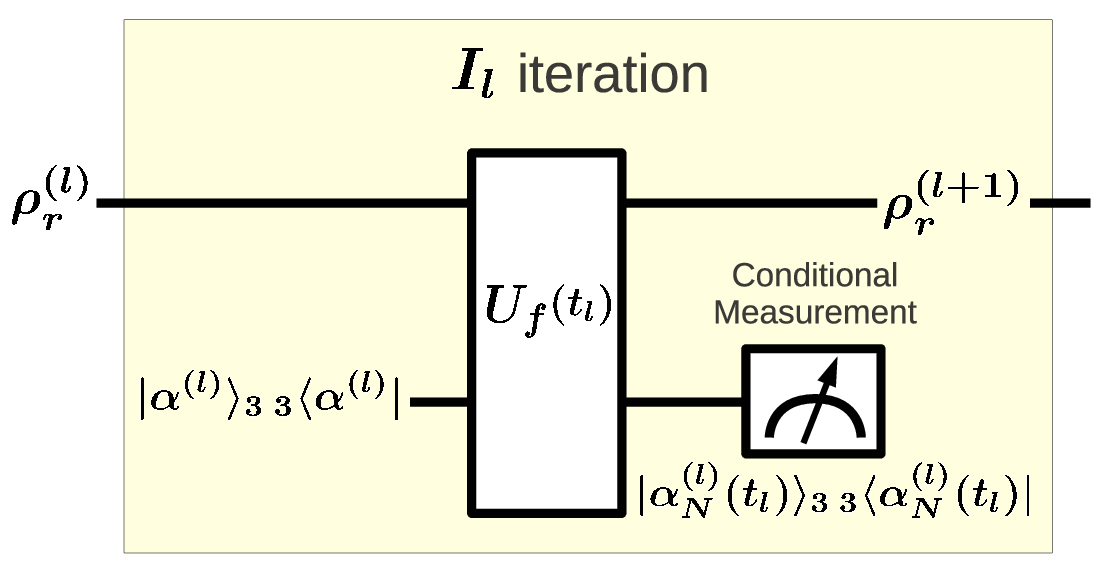}
\caption{ \label{qcir_iteration} (Color online)  The $l$-th iteration, $I_l$, in factorization.  In each iteration, 
the third harmonic oscillator is prepared in a coherent state $|\alpha^{(l)}\rangle_3$ with a magnitude $|\alpha^{(l)}|$.  
Then, a unitary operator $U_f(t_l)$ is applied.  By performing a conditional measurement of the coherent
state $|\alpha^{(l)}_N(t_l)\rangle_3$, the reduced density matrix $\rho^{(l+1)}_r$ can be obtained.
}
\end{figure}

Extending our proposal in Ref.~\cite{Ng}, we now present an amplitude-amplification method to factor any positive 
integer $N$.  
For example, let us consider three coupled harmonic oscillators for factorization by setting $A=2$ and $B=1$.
Using this method, the probability of finding the factors' state can reach nearly one after $L$ iterations.
Simplified schematic diagrams of the factorization approach are shown in 
Figs.~\ref{factqcircuit} and \ref{qcir_iteration}.

\subsection{Algorithm}
%\subsubsection{Initialization}
Now we consider the total Hamiltonian $H_f$ of the system 
\begin{equation}
\label{Hamiltonian_f}
 H_f=\hbar\sum^3_{j=1}{\omega_j}a^\dag_ja_j+{\hbar}f_{\rm factor}(n_1,n_2)a^\dag_3a_3,
\end{equation} 
where the nonlinear-operator function $f_{\rm factor}(n_1,n_2)$ for computing the product of any two trial factors 
is written as,
\begin{equation}
 f_{\rm factor}(n_1,n_2)=\sum^K_{k=1}g_k\,(n_1n_2)^k.
\end{equation}
Here the parameters $g_k$ are the coupling strengths of the nonlinear 
interactions $(n_1n_2)^ka^\dag_3a_3$, for $k=1,\ldots,K$.  
This operator function $f_{\rm factor}$ will output an eigenvalue which is a 
power series of the product $n\times{m}$, for the product state $|n,m\rangle=|n\rangle_1\;|m\rangle_2$.

The total state of the first two harmonic oscillators,  
in the number state basis, is initially prepared, i.e., 
\begin{eqnarray}
\label{initialden}
\rho^{(1)}_{r}&=&\sum_{n,n',m,m'}p^{n'm'}_{nm}|n,m\rangle\langle{n'},{m'}|,
\end{eqnarray}
where the $p^{n'm'}_{nm}$ are the probabilities of the states of 
the first and second oscillators, while $n,n',m,m'=2,\ldots,{N/2}$. 
The states of the harmonic oscillators 1 and 2 can be prepared in
arbitrary states, including pure states or mixed states \cite{Ng}.  
The states of oscillators
1 and 2 encode all trial factors of the number $N$.
Each number state $|n\rangle_j$ represents each trial factor $n$, for $j=1,2$.

\subsubsection{First iteration: $I_1$}
In the first iteration, the state of the third harmonic oscillator is prepared 
in a coherent state $|\alpha^{(1)}\rangle_3$.  
By applying the time-evolution operator $U_f(t_1)=\exp(-iH_{f}t_1)$ to 
the initial state $\rho^{(1)}(0)=\rho^{(1)}_r\otimes|\alpha^{(1)}\rangle_3\;{}_3\langle\alpha^{(1)}|$, it becomes
\begin{eqnarray}
 \rho^{(1)}(t_1)&=&U_f(t_1)\rho^{(1)}_r\otimes|\alpha^{(1)}\rangle_3\;{}_3\langle\alpha^{(1)}|U^{\dag}_f(t_1),\\
\label{rho1(t_1)}
&=&\sum_{n,m,n',m'}\tilde{p}^{n'm'}_{nm}|n,m\rangle|\alpha^{(1)}_{nm}(t_1)\rangle_3\;{}_3\langle\alpha^{(1)}_{n'm'}(t_1)|\langle{n',m'}|
\end{eqnarray}
where
\begin{equation}
\tilde{p}^{n'm'}_{nm}(t_1)=\exp\{i[(n'-n)\omega_1+(m'-m)\omega_2]t_1\}p^{n'm'}_{nm}.
\end{equation}
Here $\alpha^{(l)}_{nm}(t)$, for $l=1$, in Eq.~(\ref{rho1(t_1)}) is a complex function,
\begin{eqnarray}
\label{alpha}
\alpha^{(l)}_{nm}(t)&=&\exp(-i\Omega_{nm}t)\alpha^{(l)},
\end{eqnarray}
and $\Omega_{nm}$ is an effective rotation frequency of the coherent state in phase space
\begin{eqnarray}
\label{Theta}
\Omega_{nm}&=&\omega_3+\sum^{K}_{k=1}g_k(nm)^k.
\end{eqnarray}
Note that the product of the two factors of $r$ and $s$ is equal to $N$: $r\times{s}=N$.
The rotation frequency $\Omega_N$ of the coherent state is 
\begin{eqnarray}
\Omega_{N}&=&\omega_3+\sum^{K}_{k=1}g_kN^k.
\end{eqnarray}
If the product of any two numbers is not equal to $N$,
then the frequencies $\Omega_{nm}$ are different to the frequency $\Omega_N$. 
Thus, the state of the harmonic oscillator 3 can act as a ``marker'' for the states of factors and non-factors \cite{Ng}.

%\subsubsection{Conditional measurement}
Now we define a measurement operator $\mathcal{M}_l$ which can be written as \cite{Wiseman}
\begin{eqnarray}
\label{M_l}
\mathcal{M}_l\:\rho^{(l)}(t_l)\,&=&\,\mathcal{J}(E_{l})\rho^{(l)}(t_l)\,=\,
\sum_lE_{l}\:{\rho^{(l)}}(t_l)\:E^{\dag}_{l},\\
\label{coherent_state}
E_{l}\,&=&\,|\alpha^{(l)}_N(t_l)\rangle_3\;{}_3{\langle}\alpha^{(l)}_N(t_l)|,
\end{eqnarray}
where $\alpha^{(l)}_N(t_l)=\alpha^{(l)}\exp(-i\Omega_{N}t_l)$ and $l$ is the number of iterations.
A conditional measurement $\mathcal{M}_1$ is performed on the third oscillator 
at the time $t_1$.  
The probability of obtaining this coherent state  $|\alpha_N^{(1)}(t_1)\rangle_3$
becomes
\begin{eqnarray}
 {\rm Pr}(E_1)&=&{\rm Tr}[\mathcal{M}_1\rho^{(1)}(t_1)],\\
\label{Pr_E_1}
&=&\sum_{n,m}{p}^{nm}_{nm}\;|{\epsilon}^{(1)}_{nm}|^2,
\end{eqnarray}
where the coefficient $\epsilon^{(l)}_{nm}={}_3\langle{\alpha}^{(l)}_N(t_l)|\alpha^{(l)}_{nm}(t_l)\rangle_3$, for $l=1$, and
\begin{eqnarray}
\label{eps1}
\epsilon^{(l)}_{nm}&=&\exp\{-|\alpha^{(l)}|^2[1-\exp(i\Omega_N{t}_l-i\Omega_{nm}t_l)]\},~~~~{\rm for}~~l\geqslant{1},
\end{eqnarray}
is the overlap between the two coherent states $|\alpha^{(l)}_N(t_l)\rangle_3$ and 
$|\alpha^{(l)}_{nm}(t_l)\rangle_3$, respectively.
Note that the value of the probability ${\rm Pr}(E_1)$ can be adjusted by appropriately choosing 
the evolution time and the magnitude $|\alpha^{(1)}|$. 
In practice, this probability ${\rm Pr}(E_1)$ cannot be adjusted to be extremely small.  

Recall that a state is conditioned when this state is conditional on the measurement of a certain state \cite{Wiseman}.
After the measurement of $|\alpha^{(1)}_N(t_1)\rangle_3$, the density matrix of the conditioned state can be written as \cite{Wiseman}:
\begin{eqnarray}
\label{condst1}
 \rho^{(1)}_c&=&\frac{\mathcal{M}_1\rho^{(1)}(t_1)}{{\rm Tr}[\mathcal{M}_1\rho^{(1)}(t_1)]},\\
\label{rho^1_c}
&=&\frac{1}{C_1}\sum_{n,m,n',m'}\!\!\tilde{p}^{n'm'}_{nm}\!(t_1)\;{\epsilon}^{(1)}_{nm}\;{\epsilon}^{(1)*}_{n'm'}|n,m\rangle\langle{n',m'}|\otimes|\alpha^{(1)}_N(t_1)\rangle_3\;{}_3{\langle}\alpha^{(1)}_N(t_1)|,
\end{eqnarray}
where $C_1$ is a normalization constant
\begin{eqnarray}
\label{C_1}
 C_1&=&\sum_{n,m}{p}^{nm}_{nm}\;|{\epsilon}^{(1)}_{nm}|^2. 
\end{eqnarray}
Note that the trace of this density matrix ${\rm Tr}[\rho^{(1)}]$ is equal to one. 
After the first iteration, the probability of finding the factors of the number $N$ is 
increased by a factor $C^{-1}_1$, which is 
the inverse of the probability ${\rm Pr}(E_1)$ as seen from Eqs.~(\ref{Pr_E_1}) and (\ref{rho^1_c}).   
The probability amplification [see also Eqs.~(\ref{varrho1}) and (\ref{varrho2})] is thus inversely proportional to the 
probability of obtaining the coherent state ${\rm Pr}(E_1)$.

\subsubsection{Second iteration: $I_2$} 
After the first iteration, we now obtain the reduced density matrix $\rho^{(2)}_r$ of the first two harmonic oscillators as
\begin{eqnarray}
\label{rho2}
 \rho^{(2)}_r&=&\frac{1}{C_1}\sum_{n,m,n',m'}\tilde{p}^{n'm'}_{nm}(t_1){\epsilon}^{(1)}_{nm}{\epsilon}^{(1)*}_{n'm'}|n,m\rangle\langle{n',m'}|.
\end{eqnarray}
We consider the state $\rho^{(2)}_r$ in Eq.~(\ref{rho2}) of the oscillators 1 and 2 
as an input state for the second iteration.
The coherent state of the third harmonic oscillator is prepared in a coherent state $|\alpha^{(2)}\rangle_3$, 
with a magnitude $|\alpha^{(2)}|$.  
The nonlinear interactions between the three harmonic oscillators are then turned on for a time $t_2$.
The state evolves as 
\begin{eqnarray}
\rho^{(2)}(t_2)&=&\frac{1}{C_1}\!\sum_{n,m,n',m'}\!\!\tilde{p}^{n'm'}_{nm}\!(\tilde{t}_2)\;{\epsilon}^{(1)}_{nm}\;{\epsilon}^{(1)*}_{n'm'}|n,m\rangle|\alpha^{(2)}_{nm}(t_2)\rangle_3\;{}_3{\langle}\alpha^{(2)}_{n'm'}(t_2)|\langle{n',m'}|,
\end{eqnarray}
where $\tilde{t}_2=t_1+t_2$.

Next, a conditional measurement $\mathcal{M}_2$ is applied to the system at the time $t_2$.  
The probability of obtaining the coherent
state $|\alpha^{(2)}_N(t_2)\rangle_3$ becomes
\begin{eqnarray}
 {\rm Pr}(E_2)&=&{\rm Tr}[\mathcal{M}_2\rho^{(2)}(t_2)],\\
&=&\frac{1}{C_1}\sum_{n,m}{p}^{nm}_{nm}\;|{\epsilon}^{(1)}_{nm}|^2\;|{\epsilon}^{(2)}_{nm}|^2.
\end{eqnarray}
The conditioned state can be written as
\begin{eqnarray}
\label{condst2}
 \rho^{(2)}_c&=&\frac{\mathcal{M}_2\rho^{(2)}(t_2)}{{\rm Tr}[{\mathcal{M}_2\rho^{(2)}(t_2)}]},\\
&=&\frac{1}{C_2}\sum_{n,m,n',m'}\tilde{p}^{n'm'}_{nm}(\tilde{t}_2){\epsilon}^{(1)}_{nm}{\epsilon}^{(1)*}_{n'm'}{\epsilon}^{(2)}_{nm}{\epsilon}^{(2)*}_{n'm'}
|n,m\rangle\langle{n',m'}|\otimes|\alpha^{(2)}_N(t_2)\rangle_3\;{}_3{\langle}\alpha^{(2)}_N(t_2)|,
\end{eqnarray}
where $C_2$ is the normalization constant, 
\begin{equation}
\label{C_2}
 C_2=\sum_{n,m}{p}^{nm}_{nm}|{\epsilon}^{(1)}_{nm}|^2|{\epsilon}^{(2)}_{nm}|^2.
\end{equation}
The probability of finding the factors is enhanced by a factor $C^{-1}_2$ after the second iteration. 

The coefficients $|\epsilon^{(2)}_{nm}|$ is less than one for any product $n\times{m}\neq{N}$. 
From Eqs.~(\ref{C_1}) and (\ref{C_2}), it can be seen that $C_2<{C_1}$.  
Therefore, the probability of
finding the factors is now higher after one additional iteration.

\subsubsection{$L$-th iteration: $I_L$} 
Similarly, we now iterate the procedure $(L-1)$ times.
After $(L-1)$ iterations, the reduced density matrix of the oscillators 1 and 2 can be written as
\begin{eqnarray}
\rho^{(L)}_r&=&\frac{1}{C_{L-1}}\!\!\sum_{n,m,n',m'}\!\!\tilde{p}^{n'm'}_{nm}\!(\tilde{t}_{L-1})\prod^{L-1}_{l=1}{\epsilon}^{(l)}_{nm}\;{\epsilon}^{(l)*}_{n'm'}|n,m\rangle\langle{n',m'}|,
\end{eqnarray}
where 
\begin{eqnarray}
\label{C_{L-1}}
 C_{L-1}&=&\sum_{n,m}{p}^{nm}_{nm}\prod^{L-1}_{l=1}|{\epsilon}^{(l)}_{nm}|^2,\\
\tilde{t}_{L-1}&=&\sum^{L-1}_{l=1}t_l.
\end{eqnarray}
The state of the third harmonic oscillator is now prepared in a coherent state 
$|\alpha^{(L)}\rangle_3$, with a magnitude $|\alpha^{(L)}|$.
Let the three coupled harmonic oscillators evolve for a time $t_L$.  This gives  
\begin{eqnarray}
\rho^{(L)}(t_L)&=&\frac{1}{C_{L-1}}\!\!\sum_{n,m,n',m'}\!\!\tilde{p}^{n'm'}_{nm}\!(\tilde{t}_{L})\prod^{L-1}_{l=1}{\epsilon}^{(l)}_{nm}\;{\epsilon}^{(l)*}_{n'm'}|n,m\rangle|\alpha^{(L)}_{nm}(t_L)\rangle_3\;{}_3{\langle}\alpha^{(L)}_{n'm'}(t_L)|\langle{n',m'}|.
\end{eqnarray}
By performing a conditional measurement $|\alpha^{(L)}_N(t_L)\rangle_3$, the state becomes
\begin{eqnarray}
\label{condstL}
 \rho^{(L)}_c&=&\frac{\mathcal{M}_L\rho^{(L)}(t_L)}{{\rm Tr}[\mathcal{M}_L\rho^{(L)}(t_L)]},\\
&=&\frac{1}{C_{L}}\sum_{n,m,n',m'}\tilde{p}^{n'm'}_{nm}(\tilde{t}_L)\prod^{L}_{l=1}{\epsilon}^{(l)}_{nm}{\epsilon}^{(l)*}_{n'm'}|n,m\rangle\langle{n',m'}|\otimes|\alpha^{(L)}_N(t_L)\rangle_3\;{}_3\langle\alpha^{(L)}_N(t_L)|.
\end{eqnarray}
After the $L$-th step, the probability of finding the factors is increased by a factor $C^{-1}_L$. 
From Eq.~(\ref{C_{L-1}}), the probability of obtaining the coherent state 
$|\alpha^{(L)}_N(t_L)\rangle_3$ can be written as
\begin{eqnarray}
 {\rm Pr}(E_L)&=&\frac{1}{C_{L-1}}\sum_{n,m}{p}^{nm}_{nm}\prod^{L}_{l=1}|{\epsilon}^{(l)}_{nm}|^2,\\
\label{Pr_E_L}
&=&\frac{C_L}{C_{L-1}}.
\end{eqnarray}

The entire iterative procedure is now completed. The convergence and performance of 
this method will be discussed in the following subsections.

\subsection{Convergence}
We now study the convergence of this iterative method.
We first consider the magnitude of the coherent state $|\alpha^{(l)}|$ for each iteration as
\begin{equation}
 |\alpha^{(1)}|\leqslant|\alpha^{(2)}|\leqslant\ldots\leqslant|\alpha^{(l)}|\leqslant\ldots\leqslant|\alpha^{(L)}|.
\end{equation}
Thus, we have 
\begin{equation}
\label{decond1}
 1>|{\epsilon}^{(1)}_{nm}|^2\geqslant|{\epsilon}^{(2)}_{nm}|^2\geqslant\ldots\geqslant|{\epsilon}^{(l)}_{nm}|^2\geqslant\ldots\geqslant|{\epsilon}^{(L)}_{nm}|^2.
\end{equation}
For any product of $n$ and $m$ being not equal to $N$, 
the coefficient $|{\epsilon}^{(l)}_{nm}|^2$ is less than one and decreasing for higher $l$, and the evolution time $t_l$
is non-zero and appropriately chosen.  When the number of iterations $L$ tends to
infinity, the product of the coefficients $|{\epsilon}^{(l)}_{nm}|^2$ tends to zero,
\begin{equation}
\label{lime}
 \lim_{L\rightarrow\infty}\prod^{L}_{l=1}|{\epsilon}^{(l)}_{nm}|^2\rightarrow{0}.
\end{equation}
The coefficients $|{\epsilon}^{(l)}_{rs}|^2$ are equal to one for any product of two factors $r$ and $s$ 
being equal to $N$, i.e., when $r\times{s}=N$.

Now we consider the probability of finding a pair of factors, 
$r$ and $s$, after the $l$-th iteration, which is
\begin{eqnarray}
\label{Pr_l}
 {\rm Pr}_l\big(|{r},{s}\rangle\big)&=&\frac{p^{{r},{s}}_{{r},{s}}}{C_l}.
\end{eqnarray}
Since the coefficient $|\epsilon^{(l)}_{nm}|^2$ is less than one, the normalization constant $C_l$ in Eq.~(\ref{C_{L-1}}) 
is decreasing, i.e., $C_l\;{<}\;C_{l-1}$.  
Therefore, the probability of finding a pair of factors increases after an additional iteration.
This shows that this iterative method is convergent.

From Eqs.~(\ref{C_{L-1}}) and (\ref{lime}), it is very easy to show that 
\begin{eqnarray}
 \lim_{L\rightarrow\infty}C_{L}&=&\lim_{L\rightarrow\infty}\sum_{n,m}p^{nm}_{nm}
\prod^{L}_{l=1}|\epsilon^{(l)}_{nm}|^2\\
\label{limit_C}
&=&\sum_{r,s}{p}^{rs}_{rs}.
\end{eqnarray}
In the limit of large number of iterations $L$, we can obtain the state $\rho_f$ of the factors
\begin{eqnarray}
 \lim_{L\rightarrow\infty}\rho^{(L)}_r&\rightarrow&\rho_f,
\end{eqnarray}
where 
\begin{eqnarray}
\rho_f=\frac{1}{C^*_f}\sum_{r,s,r',s'}{p}^{r's'}_{rs}|r,s\rangle\langle{r',s'}|,
\end{eqnarray}
and $C^*_f=\sum_{r,s}{p}^{rs}_{rs}$, with $r,s,r',s'$ are factors of $N$ ($r{\times}s=r'{\times}s'=N$). 
This shows that the state of the factors can be achieved by employing this iterative method, if a
sufficiently large number of iterations is used.

\subsection{Performance}
We can now estimate the number $L$ of the iterations required to achieve a probability of order of one for 
factoring $N$. 
We investigate the amplification ratio $\lambda_l$ of the two probabilities of finding the factors 
${\rm Pr}_{l-1}\big(|{r},{s}\rangle\big)$ and ${\rm Pr}_l\big(|{r},{s}\rangle\big)$ after the $(l-1)$-th and the $l$-th iterations.
From Eq.~(\ref{Pr_l}), we have 
\begin{eqnarray}
\label{varrho1}
\lambda_l&=&\frac{{\rm Pr}_{l}\big(|{r},{s}\rangle\big)}{{\rm Pr}_{l-1}\big(|{r},{s}\rangle\big)},\\
\label{varrho2}
&=&\frac{C_{l-1}}{C_{l}}.
\end{eqnarray}
Note that this ratio $\lambda_l$ is just the reciprocal ${\rm Pr}^{-1}(E_l)$
in Eq.~(\ref{Pr_E_L}) of the probability of obtaining the coherent state.
Practically, this probability ${\rm Pr}(E_l)$ cannot be too small. 
For example, let the amplification ratio $\lambda_l$ be roughly equal to 
${\lambda}{~\sim~}O(10^{-1})$ for each iteration, and let the probability 
of the factors' state ${\rm Pr}_0\big(|r,s\rangle\big)$ before the iterations be
of order ${O}(N^{-z})$, where $z$ is a positive number.  
After $L$ iterations, the probability of
finding the states of the factors can be increased by a factor $\lambda^{L}$.
Here we require
\begin{eqnarray}
\label{Lcond}
 {O}(\lambda^{L})\sim{O}(N^z).
\end{eqnarray}
Therefore, we obtain that the number $L$ of necessary iterations is of order of ${O}({\log_2{N}})$.   
In the limit of large $L$, the probabilities ${\rm Pr}(E_l)$ in Eq.~(\ref{Pr_E_L}) tend to one because $C_L$ 
approaches the sum of probabilities of the factors' states  in Eq.~(\ref{limit_C}).
The probability of finding the factors will be slowly increased after the
number $L\sim{O}(\log_2{N})$ of iterations is reached.

\subsection{Example: Initial pure states}
In this section, we study how to factorize an integer $N$ with an initial pure state
using this factoring algorithm.   We consider the initial state of the first two harmonic oscillators
as the superposition of number states, i.e., 
\begin{eqnarray}
\label{initial_pure}
|\Psi(0)\rangle&=&\Bigg(\frac{1}{D_1}\sum^{{\lceil}\sqrt{N}\!~{\rceil}}_{n=3}|n\rangle_1\Bigg)\Bigg(\frac{1}{D_2}\sum^{{\lceil}{N/3}{\rceil}}_{m=\lceil\sqrt{N+1}\!~\rceil}|m\rangle_2\Bigg),
\end{eqnarray}
where 
\begin{eqnarray}
D_1&=&({\big{\lceil}\sqrt{N}\!~\big{\rceil}-2})^{1/2}, \\
D_2&=&({\big{\lceil}{N/3}\big{\rceil}-\lceil\sqrt{N+1}\!~\rceil}+1)^{1/2},
\end{eqnarray}
are two normalization constants.  Here we consider trial factors from 3, $\ldots$, $N/3$. 
The probability of finding the product of two factors is of order of ${O}(N^{-3/2})$.  

For example, now we show how to factor the integer $N=1,030,189=1009{\times}1021$. 
For simplicity, we now take $K=1$, which is the lowest order of nonlinearity.  The Hamiltonian can be 
written as
\begin{eqnarray}
\label{Hamiltonian_1}
 H_1=\hbar\sum^3_{j=1}{\omega_j}a^\dag_ja_j+{\hbar}{g}a^\dag_1a_1a^\dag_2a_2a^\dag_3a_3,
\end{eqnarray}
where $g$ is the nonlinear strength.  The stronger nonlinear strengths and 
high-order nonlinearity can significantly 
shorten the required time evolution of the system \cite{Ng}.  But the role of 
nonlinearity is not directly relevant 
to the number $L$ of the required iterations for the amplitude amplification.  
\begin{table}[ht]
\caption{\label{table1}
This table shows the fidelities $F_l$, the probabilities Pr$(E_l)$ of obtaining 
the coherent states $|\alpha^{(l)}_N(t^*_l)\rangle_3$ in Eq.~(\ref{coherent_state}), and the evolution time $t^*_l$ for the $l$ iterations.
Here $E_l$ is the density matrix of the coherent state $|\alpha^{(l)}_N(t^*_l)\rangle_3$ in Eq.~(\ref{coherent_state}) and $t^*_l$
is measured in units of $g^{-1}$.}
 \begin{ruledtabular}
  \begin{tabular}{cccccccc}
\hline
   Iterations & Fidelities & Probabilities for coherent states  & Evolution times                                    \\
   $l$ & $F_l$ & Pr$(E_l)$  & $t^*_l$ \\ 
\hline
 1 & 2.010$\times10^{-8}$ & 0.143 & 1.704 \\
 2 & 1.403$\times10^{-7}$ & 0.143 & 1.342 \\
 3 & 9.782$\times10^{-7}$ & 0.143 & 5.000  \\
 4 & 6.821$\times10^{-6}$ & 0.143 & 4.610 \\
 5 & 4.739$\times10^{-5}$ & 0.144 & 0.732  \\
 6 & 3.259$\times10^{-4}$ & 0.145 & 3.108 \\
 7 & 2.172$\times10^{-3}$ & 0.150 & 1.635 \\
 8 & 1.445$\times10^{-2}$ & 0.150 & 4.559  \\
 9 & 1.045$\times10^{-1}$ & 0.138 & 4.222  \\
 10 & 5.092$\times10^{-1}$ & 0.205 & 6.046 \\
 11 & 8.506$\times10^{-1}$ & 0.599 & 2.434  \\
 12 & 9.919$\times10^{-1}$ & 0.858 & 1.175  \\
 13 & 9.985$\times10^{-1}$ & 0.994 & 5.089  \\
 14 & 9.997$\times10^{-1}$ & 0.999 & 5.833  \\
 15 & 1.000$\times10^{-1}$ & 1.000 & 0.708 \\
\hline
  \end{tabular}
\end{ruledtabular}
\end{table}

We take $t^*_l$ as the evolution time for the $l$-th iteration,
\begin{eqnarray}
\label{tr}
 t^*_l&=&\frac{2\pi}{g}r_l,
\end{eqnarray}
where $r_l$ is a uniformly distributed random number on the interval $[0,1]$. 
We now evaluate the performance of this method by investigating the fidelity $F_l$
between the reduced density matrix $\rho^{(l+1)}_r$ and the factor's states $\rho_f$
as \cite{Uhlmann,Jozsa}
\begin{eqnarray}
 F_l&=&\Big\{{\rm Tr}\Big[\big({{\rho^{1/2}_f}\rho^{(l+1)}_r{\rho^{1/2}_f}}\big)^{1/2}\Big]\Big\}^2.
\end{eqnarray}

\begin{figure}[ht]
\centering
\includegraphics[height=14.5cm]{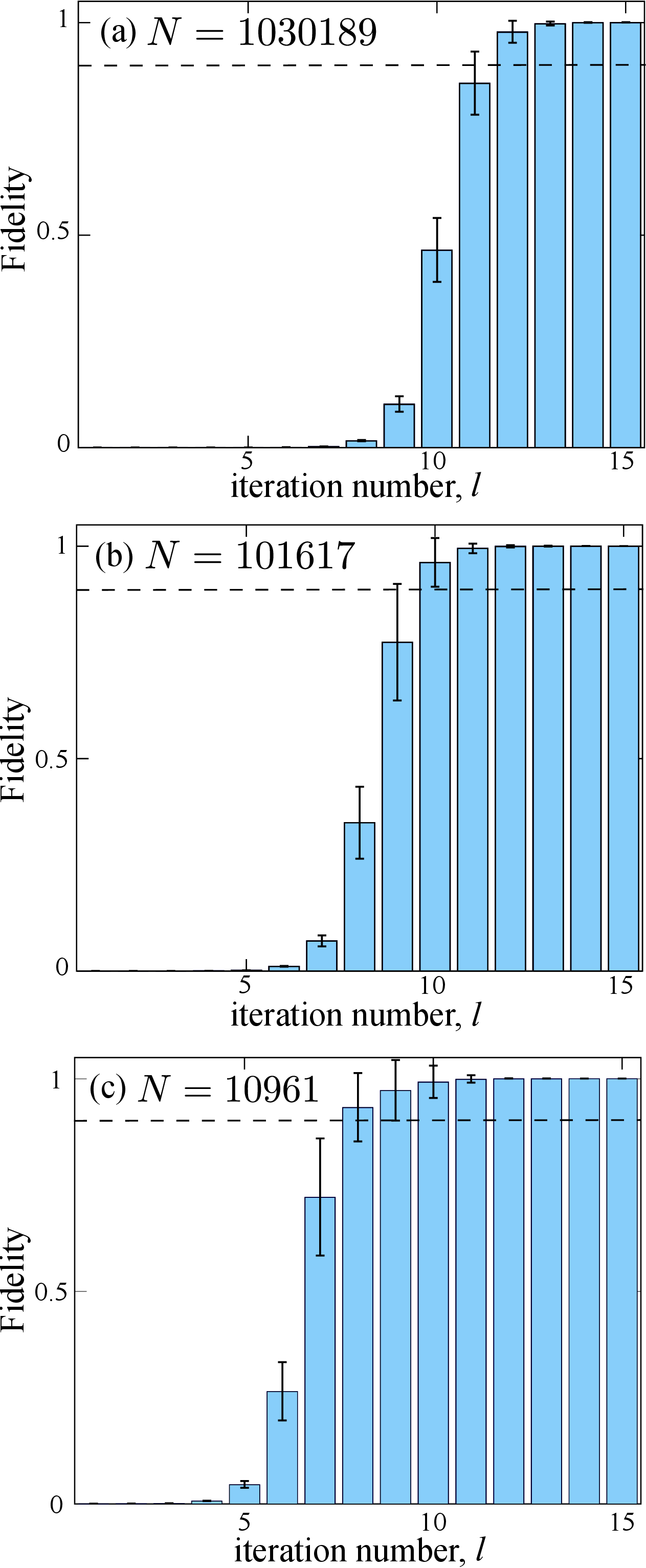}
\caption{ \label{figfid} (Color online) Bar chart: the average of the fidelities between the reduced density matrix $\rho^{(l)}_r$
and the factors' states $\rho_f$ are plotted versus the number $l$ of iterations in (a) $N=1,030,189$, (b) $N=101,617$ and (c) $N=10,961$.  
In each iteration, the fidelity $F_l$ is taken at the time $t^*_l$, which is a uniformly distributed random number ranging
from 0 to $2\pi/g$.  Here the sample size is equal to 100 fidelities.
The bars, which indicate the mean values, are shown in light blue.  The error bars, which indicate the standard deviations, 
are shown in black. The dashed lines indicate that the
fidelities attain 0.9.}
\end{figure}

Table \ref{table1} shows the relevant fidelity, the probability of obtaining the coherent 
state $|\alpha^{(l)}_N(t^*_l)\rangle_3$ with a magnitude $|\alpha^{(l)}|=2$, and the evolution time $t^*_l$ 
for each iteration. 
Before starting the iteration, the initial fidelity $F_0$ is very low: $2.883\times10^{-9}$.
In the first few steps, the probabilities, Pr$(E_l)$, of obtaining the coherent states, 
are about $10^{-1}$.  We emphasize that now the probabilities Pr$(E_l)$ are not
extremely small, even when $N$ is large.  This resolves the limitation
of our previous proposal \cite{Ng}.
The probability of finding the factor's state can be increased by 10 after a single iteration.
After ten iterations, the fidelity can exceed 0.5.  
The fidelity can reach nearly one after two more iterations, while the probabilities $P(E_l)$ 
approach one.  Also, we have explicitly shown that this method can factor
a number of order of $10^6$ with 12 iterations.

Since the evolution time $t^*_l$ in Eq.~(\ref{tr}) is a uniformly distributed random number 
from 0 to $2\pi/g$, it is necessary to study the effect of this randomly chosen time $t^*_l$
to the performance of factoring.
Now we study the averages 
of the fidelities $F_l$ and probabilities Pr$(E_l)$ of obtaining the coherent states.  
We also examine their standard deviations in each iteration.

In Fig.~\ref{figfid}(a), we plot the average of the fidelities $F_l$
versus the number $l$ of iterations.  Here we take the sample size to be equal to 
100 fidelities in each iteration.  As shown in Fig.~\ref{figfid}, the average of the fidelities is about 0.5 at the 10-th iteration.
The fidelity is greater than 0.9 at the 12-th iteration.
After 12 iterations, the mean values of the fidelities reach nearly one.  
The error bars, which indicate
the standard deviations, are shown in the same figure.  
We can see that, after the 12-$th$ iteration, the standard deviations are relatively small compared to the mean values
in each iteration.

We then plot the average of the fidelities $F_l$
versus the number $l$ of iterations for $N=10,1617$
and 10,961 in Fig.~\ref{figfid}(a) and (b), respectively.
The fidelities exceed 0.9 after the 10-th iteration for 
$N=101,617$ and the 8-th iteration for $N=10,961$.
The numerical results in Fig.~\ref{figfid} show that 
the required number of iterations increases logarithmically 
with $N$ for which the fidelity is greater than 0.9.

\begin{figure}[ht]
\centering
\includegraphics[height=14.5cm]{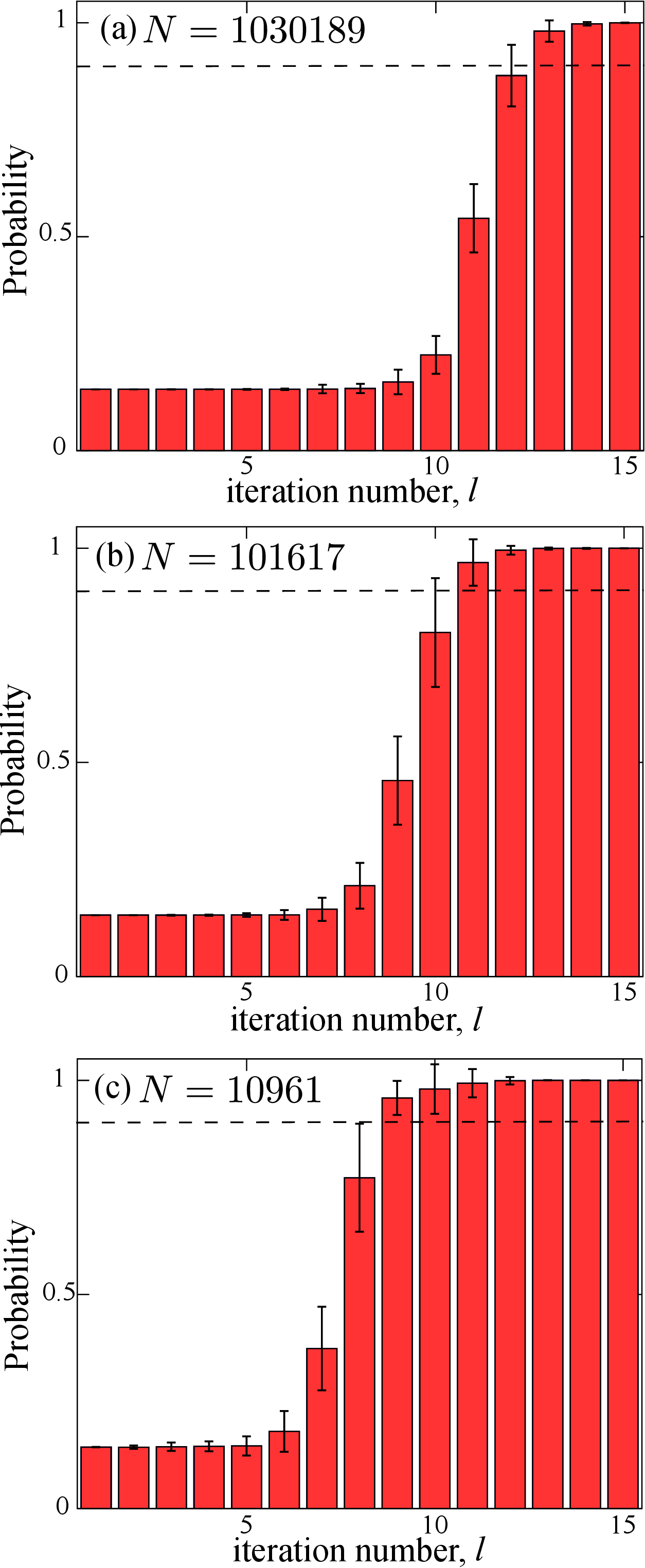}
\caption{ \label{figpb} (Color online) Bar chart: the average of the probabilities of 
obtaining the coherent states $|\alpha^{(l)}_N(t^*_l)\rangle_3$ 
are plotted versus the number $l$ of iterations in (a) $N=1,030,189$,
(b) $N=101,617$ and (c) $N=10,961$.
In each iteration, the probability is taken at the 
time $t^*_l$ which is a uniformly distributed random number from 0 to $2\pi/g$.
Here the sample size is taken to be 100 probabilities.
The bars, which indicate the mean values, are in red.  The error bars, which indicate the standard deviations, 
are shown in black. The dashed lines indicate that the
probabilities attain 0.9.
}
\end{figure}

In Fig.~\ref{figpb}(a), the average of the probabilities Pr$(E_l)$ of obtaining the coherent states 
is plotted versus the number $l$ of iterations.  In the first ten iterations, 
the probabilities Pr$(E_l)$ are about 0.1.  Then, it increases and 
saturates around one after 13 iterations.
Moreover, these standard deviations are much smaller than the mean values of Pr$(E_l)$.
This means that the statistical effect of the time $t^*_l$ is small on the performance
of quantum factorization.

In Fig.~\ref{figpb}(b) and (c), the average of the probabilities Pr$(E_l)$ of obtaining the coherent states are plotted versus the number $l$ of iterations for $N=101,617$ and $N=10,961$, respectively.
The probabilities are greater than 0.9 after the 11-th iteration for 
$N=101,617$ and the 9-th iteration for $N=10,961$.
Therefore, the results show that the number of iterations (Pr$(E_l){\geq}0.9$) also logarithmically scales with $N$.

\newpage
\section{Limitations and problems}
We have introduced an amplitude-amplification method for factoring 
integers by using three coupled harmonic oscillators. However, 
this method requires an exponential amount of resources 
for implementation.  
This means that the required 
resource is of order 
of O$({N})$, where $N$ is the input size.  
This means that this approach cannot provide any speed-up
compared to classical algorithms. We discuss 
these two limitations in the following subsections.

\subsection{Exponential energy resource}
First, this approach requires an exponential amount of energy resource to 
encode a number $N$ onto the state of a single harmonic oscillator.
For example, to factorize a number $N$, two harmonic 
oscillators are used for encoding $O(N^{3/2})$ possible states of 
trial factors. The energy $E~\sim~\hbar{\omega}O(N^{3/2})$
is thus required, where $\omega\approx{\omega_{1,2}}$ 
are the frequencies of the harmonic oscillators 1 and 2.  
The required energy scales with the size of the input number
$N$. Therefore, {\it the energy becomes enormous when the 
encoded number is large.}  This encoding method becomes impractical 
when a large input number is used. This problem may be 
resolved by using another encoding method.
For example, the number could be encoded either onto qubit or
qudit states.  Thus, the required energy for encoding numbers
could be reduced.

\subsection{Exponential size of the ensemble}
Second, this method requires an exponential size of
the ensemble of harmonic oscillators for an input $N$.  
In each iteration, it is 
necessary to abandon a large 
number of harmonic oscillators after each conditional
measurement.  The number of abandoned oscillators
is proportional to the failure probability of obtaining conditional
measurement of the coherent states in each step.  
Therefore, a number at least of order of O$({N})$ 
harmonic oscillators are needed in order 
to complete the entire procedure. 
This requires an exponential resource for preparing 
the ensemble of oscillators.  This approach
may be improved by employing efficient methods
for preparing the ensembles.

\section{Conclusions}
We have presented an amplitude-amplification method
by repeatedly using the nonlinear interactions
between the harmonic oscillators and non-orthogonal measurements. 
We have shown that this approach can 
be used for factoring integer, and the factors of 
an integer $N$ can be obtained, with a high probability, 
by using a number of iterations $L\sim{O(\log_2{N})}$. 
We have numerically studied an example for factoring
$N=1,030,189$, respectively.
We have shown how to factorize an integer of order
of $O(10^6)$ within 12 iterations. 
In each iteration, the probability of obtaining this 
coherent state, with the rotation frequency
being proportional to $N$, is not less than 0.1.
By comparing examples with $N=101,617$ and $N=10,961$,
we have shown that the required
number of iterations 
increases logarithmically with $N$. 

However, using coupled harmonic oscillators,
this method requires the use of an exponential amount 
of resources, i.e., exponential energy and 
ensemble size.  Thus, this approach becomes impractical for
large input sizes.  We hope that the 
resolutions could be proposed to overcome
the problems of this approach.

Also, we stress that the nonlinear interactions between the coupled oscillators and conditional measurements are essential in this approach.
By appropriately controlling nonlinear interactions between
the coupled harmonic oscillators, the functions with integer inputs can be evaluated in a single operation.
To implement this approach, it is necessary to engineer ``many-body''
interactions of the system of harmonic oscillators.  For example, to perform quantum factorization,
it is required to generate ``three-body'' interactions between the harmonic oscillators.  We have
briefly discussed the possible implementations in Ref.~\cite{Ng}.  One of the promising candidates is neutral atoms
or polar molecules trapped in optical lattices \cite{Buchler,Johnson,Will}.  The ``three-body'' interactions can be tuned
by external fields \cite{Buchler,Johnson}.

\begin{acknowledgments}
FN acknowledges partial support from the 
Army Research Office, 
JSPS-RFBR contract No.~09-02-92114, 
Grant-in-Aid for Scientific Research (S), 
MEXT Kakenhi on Quantum Cybernetics, and the
Funding Program for Innovative R\&D on Science 
and Technology (FIRST).
\end{acknowledgments}

\end{document}